\documentclass[preprint,showpacs,preprintnumbers,amsmath,amssymb]{revtex4}

\usepackage{graphicx}
\usepackage{dcolumn}
\usepackage{bm}

\begin{document}

\title{Moment Equations for a Spatially Extended System of Two Competing Species}
\author{\large D. Valenti$^a$\footnote{e-mail: valentid@gip.dft.unipa.it},
L. Schimansky-Geier$^b$, X. Sailer$^b$, and B. Spagnolo$^a$}
\affiliation{$^a$Dipartimento di Fisica e Tecnologie Relative,
Universit\`a di Palermo\\ and INFM-CNR, Group of Interdisciplinary
Physics\footnote{Electronic address: http://gip.dft.unipa.it},
Viale delle Scienze, I-90128 Palermo, Italy\smallskip \\
$^b$Institut f\"{u}r Physik, Humboldt Universit\"{a}t zu
Berlin\\
Newtonstra{\ss}e 15, D-124891 Berlin, Germany}

\date{\today}

\begin{abstract}
The dynamics of a spatially extended system of two competing
species in the presence of two noise sources is studied. A
correlated dichotomous noise acts on the interaction parameter and
a multiplicative white noise affects directly the dynamics of the
two species. To describe the spatial distribution of the species
we use a model based on Lotka-Volterra (LV) equations. By writing
them in a mean field form, the corresponding moment equations for
the species concentrations are obtained in Gaussian approximation.
In this formalism the system dynamics is analyzed for different
values of the multiplicative noise intensity. Finally by comparing
these results with those obtained by direct simulations of the
time discrete version of LV equations, that is coupled map lattice
(CML) model, we conclude that the anticorrelated oscillations of
the species densities are strictly related to non-overlapping
spatial patterns.
\end{abstract}




\maketitle



\section{Introduction}
\vskip-0.2cm The dynamics of real ecosystems is strongly affected by
the presence of noise sources, such as the random variability of
temperature, resources and in general environment, with which the
system has a multiplicative interaction~\cite{Zimmer,Ciuchi}. In
this paper we analyze the time evolution of a spatially extended
system formed by two competing species in the presence of two noise
sources. We get the dynamics in the formalism of the moments. We
study the role of the two noise sources on the ecosystem dynamics,
described by generalized Lotka-Volterra equations in the presence of
external fluctuations, modelled as multiplicative noise.
Specifically we focus on the time behavior of the $1^{st}$ and
$2^{nd}$ order moments of the species concentrations. We find that
the $1^{st}$ order moments are independent on the multiplicative
noise intensity. On the other hand the behavior of the $2^{nd}$
order moments is strongly affected by the presence of a source of
external noise. We find anticorrelated time behavior of the species
densities. Comparing our results with those obtained by calculating
the same quantities within a coupled map lattice (CML)
model~\cite{Kaneko}, we conclude that the anticorrelated
oscillations of the species concentrations are strictly related to
non-overlapping spatial patterns~\cite{Spagnolo}. Our theoretical
results could match data from a real ecosystem, whose dynamics is
affected by the random variability of the environment, and could
provide useful tools to predict behavior of biological
species~\cite{Zimmer,Ciuchi,Spagnolo,Garcia}.
\section{The model}
\vskip-0.2cm Our system is described by a time evolution model of
Lotka-Volterra equations, within the Ito scheme, with diffusive
terms in a spatial lattice with $N$ sites
\begin{eqnarray}
\dot{x}_{i,j}&=&\mu x_{i,j} (1-x_{i,j}-\beta y_{i,j})+x_{i,j}
\sqrt{\sigma_x}\xi^x_{i,j} + D\sum_\gamma (x_\gamma-x_{i,j})
\label{Lotka_eq_1}\\
\dot{y}_{i,j}&=&\mu y_{i,j} (1-y_{i,j}-\beta x_{i,j})+ y_{i,j}
\sqrt{\sigma_y}\xi^y_{i,j} + D\sum_\gamma (y_{\gamma}-y_{i,j}),
\label{Lotka_eq_2}
\end{eqnarray}
where $x_{i,j}$ and $y_{i,j}$ denote respectively the densities of
species $x$ and species $y$ in the lattice site $(i,j)$, $\mu$ is
the growth rate, $D$ is the diffusion constant, and $\sum_\gamma$
indicates the sum over all the sites. Here $\xi^x_{i,j}(t)$ and
$\xi^y_{i,j}(t)$ are statistically independent Gaussian white
noises with zero mean and unit variance, $\sigma_x$ and $\sigma_y$
are the intensities of the multiplicative noise which models the
interaction between the species and the environment, and $\beta$
is the interaction parameter.

\subsection{The interaction parameter}
\begin{figure}[htbp]
\begin{center}
\includegraphics[width=15cm]{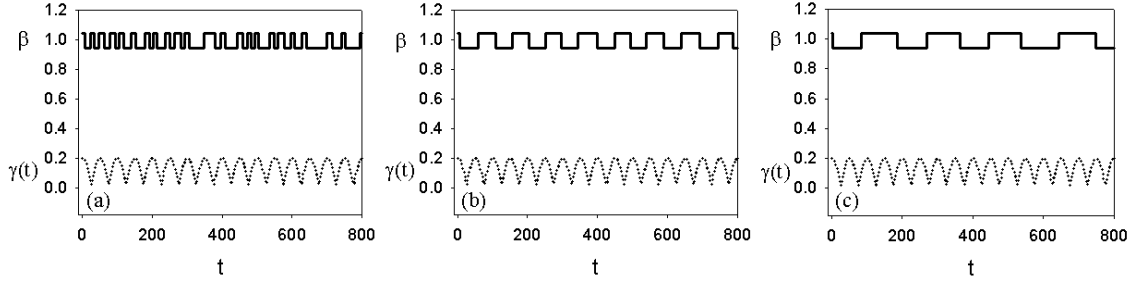}
\end{center}
\vskip-0.6cm \caption{\small Time evolution of the interaction
parameter $\beta(t)$ with initial value $\beta(0)=1.04$ and
different values of delay: (a) $\tau_d = 10$ (a), $43.5$ (b), and
$80$ (c). The values of the other parameters are: $A =9.0$,
$\omega/(2\pi)=10^{-2}$, $\gamma_0=2\cdot10^{-2}$.\bigskip}
 \label{beta}
 \vskip-0.3cm
\end{figure}

\vskip-0.2cm Depending on the value of the interaction parameter,
coexistence or exclusion regimes take place. Namely for $\beta < 1$
both species survives, while for $\beta > 1$ one of the two species
extinguishes after a certain time. These two regimes correspond to
stable states of the Lotka-Volterra's deterministic
model~\cite{Spagnolo,Vilar-Spagnolo,Valenti,Valenti1}. Moreover
periodical and random driving forces connected with environmental
and climatic variables, such as the temperature, modify the dynamics
of the ecosystem, affecting both directly the species densities and
the interaction parameter. This causes the system dynamics to change
between coexistence ($\beta < 1$) and exclusion ($\beta > 1$)
regimes. To describe this dynamical behavior we consider as
interaction parameter $\beta(t)$ a dichotomous stochastic process,
whose jump rate is a periodic function $\gamma(t)$
\begin{align}
\gamma(t) = \left\{
\begin{array}
[c]{ll}%
0, &\quad  \Delta t \leq \tau_d\\
\gamma_0 \left(1 + A \thinspace \vert \cos\omega t \vert \right),
&\quad \Delta t > \tau_d\;.
\end{array}
\right.
\label{jump_rate}
\end{align}
Here $\Delta t$ is the time interval between two consecutive
switches, and $\tau_d$ is the delay between two jumps, that is the
time interval after a switch, before another jump can occur. In
eq.~(\ref{jump_rate}), $A$ and $\omega = (2\pi)/T$ are
respectively the amplitude and the angular frequency of the
periodic term, and $\gamma_0$ is the jump rate in the absence of
periodic term. This causes $\beta(t)$ to jump between two values,
$\beta_{down} < 1$ and $\beta_{up} > 1$, which correspond to the
dynamical regimes of the deterministic Lotka-Volterra's model
(coexistence and exclusion regions). Because the dynamics of the
species strongly depends on the value of the interaction
parameter, we report in Fig.~\ref{beta} the time series of
$\beta(t)$ for different values of delay $\tau_d$, namely $\tau_d
= 10, 43.5, 60$, with $\beta_{down} = 0.94$ and $\beta_{up} =
1.04$.
We note that the correlation time $\tau_d$ of the dichotomous noise
affects the switch time between the two levels of $\beta(t)$. For a
delay time a bit less than $T/2$, we observe a synchronization
between the jumps and the periodicity of the rate $\gamma(t)$. This
synchronization phenomenon is due to the choice of the $\tau_d$
value, which stabilizes the jumps in such a way they happen for high
values of the jump rate, that is for values around the maximum of
the function $\gamma(t)$. This causes a quasi-periodical time
behavior of the species concentrations $x$ and $y$, which can be
considered as a signature of the stochastic resonance
phenomenon~\cite{Benzi} in population
dynamics~\cite{Vilar-Spagnolo,Valenti,Valenti1}. Therefore we fix
the delay at the value $\tau_D=43.5$, corresponding to a competition
regime with $\beta$ switching quasi-periodically from coexistence to
exclusion regions (see Fig.~\ref{beta}b).
\section{Mean field model}
\vskip-0.2cm In this section we derive the moment equations for our
system. Assuming $N \rightarrow \infty$, we write
Eqs.~(\ref{Lotka_eq_1}) and (\ref{Lotka_eq_2}) in a mean field form
\begin{eqnarray}
\dot{x}&=&f_x(x,y)+\sqrt{\sigma_x} g_x(x) \xi^x + D(<x>-x),
\label{mean_eq_1}\\
\dot{y}&=&f_y(x,y)+\sqrt{\sigma_y} g_y(y) \xi^y + D(<y>-y),
\label{mean_eq_2}
\end{eqnarray}
where $<x>$ and $<y>$ are average values on the spatial lattice
considered (strictly speaking they are the ensemble average in the
thermodynamics limit) and we set $f_x(x,y)=\mu x (1-x-\beta y)$,
$g_x(x)=x$, $f_y(x,y)=\mu y (1-y-\beta x)$, $g_y(y)=y$. By site
averaging Eqs.~(\ref{mean_eq_1}) and (\ref{mean_eq_2}), we obtain
\begin{eqnarray}
<\dot{x}>~=~<f_x(x,y)>,
\label{mean site eq1}\\
<\dot{y}>~=~<f_y(x,y)>.
\label{mean site eq2}
\end{eqnarray}
By expanding the functions $f_x(x,y)$, $g_x(x)$, $f_y(x,y)$,
$g_y(y)$ around the $1^{st}$ order moments $<x>$ and $<y>$, we get
an infinite set of simultaneous ordinary differential equations
for all the moments~\cite{Kawai}. To truncate this set we apply a
Gaussian approximation, for which the cumulants above the $2^{nd}$
order vanish. Therefore we obtain
\begin{eqnarray}
<\dot{x}>&=&\mu <x> (1-<x>-\beta <y>) - \mu(\beta\mu_{11}+\mu_{20})
\label{x_mean_eq}\\
<\dot{y}>&=&\mu <y> (1-<y>-\beta <x>) -
\mu(\beta\mu_{11}+\mu_{02}) \label{y_mean_eq}\\
\dot{\mu}_{20}&=&2\mu \mu_{20} - 2D\mu_{20} - 2\mu\beta<y>\mu_{20}
- 2\mu <x>(\beta\mu_{11}+2\mu_{20})\nonumber\\
&+& 2\sigma_x(<x>^2+\mu_{20})\label{mu20_eq}\\
\dot{\mu}_{02}&=&2\mu \mu_{02} - 2D\mu_{02} - 2\mu\beta<x>\mu_{02}
- 2\mu <y>(\beta\mu_{11}+2\mu_{02})\nonumber\\
&+& 2\sigma_y(<y>^2+\mu_{02})\label{mu02_eq}\\
\dot{\mu}_{11}&=&2\mu\mu_{11} - 2D\mu_{11}
-<x>[2\mu\mu_{11}+\mu\beta(\mu_{11}+\mu_{02})]\nonumber\\
&-&<y>[2\mu\mu_{11}+\mu\beta(\mu_{11}+\mu_{20})]~,
 \label{mu11_eq}
\end{eqnarray}
where $\mu_{20}$, $\mu_{02}$, $\mu_{11}$ are the $2^{nd}$ order
central moments defined on the lattice
\begin{eqnarray}
\mu_{20}&=&<x^2>-<x>^2\label{mu20}\\
\mu_{02}&=&<y^2>-<y>^2\label{mu02}\\
\mu_{11}&=&<xy>-<x><y>~.\label{mu11}
\end{eqnarray}
In order to get the dynamics of the two species we analyze the
time evolution of the $1^{st}$ and $2^{nd}$ order moments
according to Eqs.~(\ref{x_mean_eq})-(\ref{mu11_eq}). We fix the
delay time at the value $\tau_d = 43.5$, corresponding to a
quasi-periodic switching between the coexistence and exclusion
regimes, and we obtain the time series of the moments for two
values of the multiplicative noise intensity
$\sigma=\sigma_x=\sigma_y$, namely $\sigma = 10^{-12}$, $10^{-6}$,
and in the absence of it. The values of the parameters are
$\mu=2$, $D=0.05$. The initial values of the moments are $<x(0)>$
= $<y(0)>$ = $0.1$, $\mu_{20}(0)= \mu_{02}(0) = \mu_{11}(0)= 0$.
These initial conditions correspond to uniformly distributed
species on the lattice considered.
\begin{figure}[htbp]
\begin{center}
\includegraphics[width=12cm]{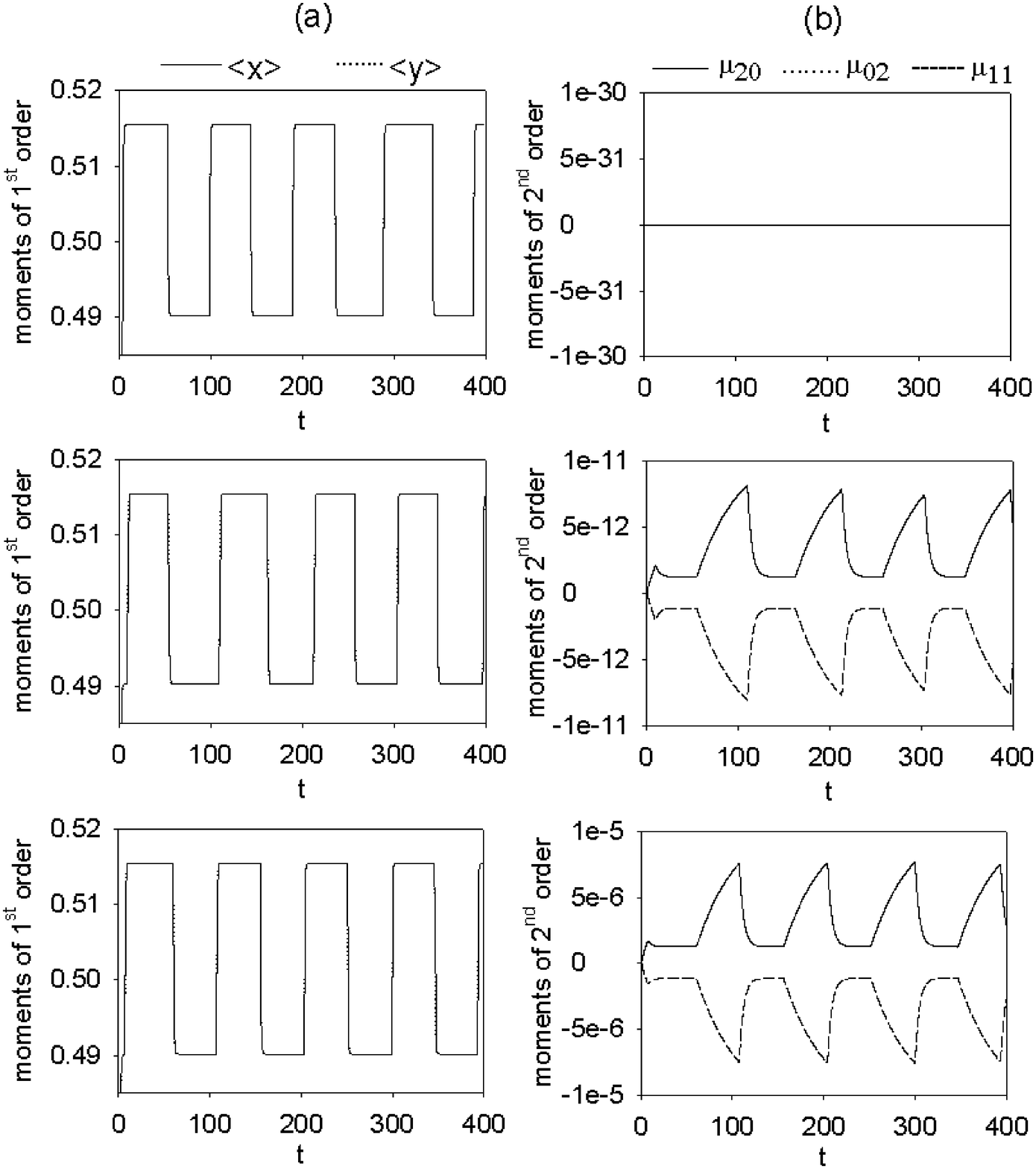}
\end{center}
\vskip-0.6cm \caption{\small Time evolution of the $1^{st}$ and
$2^{nd}$ order moments. The time series of (a) $<x(t)>$ and
$<y(t)>$, and (b) $\mu_{2,0}$ and $\mu_{0,2}$ respectively, are
completely overlapped. The values of the multiplicative noise
intensity are: $\sigma=0, 10^{-12}, 10^{-6}$, from top to bottom.
Here $\tau_d = 43.5$, $\mu=2$, and $D=0.05$. The initial values of
the moments are $<x(0)>$ = $<y(0)>$ = $0.1$,
$\mu_{20}(0)=\mu_{02}(0)=\mu_{11}(0)=0$. The values of the other
parameters are the same of Fig.~\ref{beta}.\bigskip}
 \label{moment_series}
 \vskip-0.3cm
\end{figure}
In Fig.~\ref{moment_series} we note that the $1^{st}$ order
moments of both species oscillate together quasi regularly around
$0.5$, independently on the multiplicative noise intensity (see
Fig.~\ref{moment_series}a). The noise intensity affects strongly
the dynamics of the $2^{nd}$ order moments. In the absence of
noise $\mu_{20}$, $\mu_{02}$, $\mu_{11}$ are zero. For very low
levels of multiplicative noise ($\sigma=10^{-12}$)
quasi-periodical oscillations appear with the same frequency of
the interaction parameter $\beta (t)$, because the noise breaks
the symmetry of the dynamical behavior of the $2^{nd}$ order
moments (see Fig.~\ref{moment_series}b)~\cite{Valenti}. The time
behavior of the variances of x and y species coincides all the
time with alternating periods, characterized by small (close to
zero) and large values. However the negative values of the
correlation $\mu_{11}$ indicate that the two species distributions
are anti-correlated. This means that the spatial distribution in
the lattice will be characterized by zones with a maximum of
concentration of species $x$ and a minimum of concentration of
species $y$ and viceversa. The two species will be distributed
therefore in non-overlapping spatial patterns. This physical
picture is in agreement with previous results obtained with a
different model ~\cite{Valenti1}. For higher levels of
multiplicative noise ($\sigma = 10^{-6}$) the amplitude of the
oscillations increases both in $\mu_{20}$, $\mu_{02}$ and
$\mu_{11}$. This gives information on the probability density of
both species, whose width and mean value undergo the same
oscillating behavior. The anti-correlated behavior is enhanced by
increasing the noise intensity value (see
Figs.~\ref{moment_series}b). We note that the amplitude of the
oscillations in Fig.~\ref{moment_series}b increases with the noise
intensity $\sigma$ and it is of the same order of magnitude. The
periodicity of these noise-induced oscillations shown in
Fig.~\ref{moment_series} is the same of the interaction parameter
$\beta(t)$ (see Fig.~\ref{beta}). Even if it is due to a very
different mechanism, this behavior is similar to the stochastic
resonance effect produced in population dynamics, when the
interaction parameter is subjected to an oscillating bistable
potential in the presence of additive
noise~\cite{Valenti,Valenti1}. We note that in the absence of
external noise ($\sigma = 0$) both populations coexist and the
species densities oscillate in phase around their stationary
value~\cite{Valenti}. This occurs identically in each site of the
spatial lattice. The behavior of the mean value therefore will
reproduce this situation. For $\sigma \neq 0$, anticorrelated
oscillations appear due to the multiplicative noise, superimposed
to the average behavior obtained for $\sigma = 0$ and distributed
randomly in the spatial structure. By site averaging these
noise-induced oscillations (see Ref.~\cite{Valenti}) we recover
the average behavior obtained in the absence of noise. This
explains why the first moment behavior is independent on the
external noise intensity.

\section{Coupled Map Lattice Model}
\begin{figure}[htbp]
\begin{center}
\includegraphics[width=12cm]{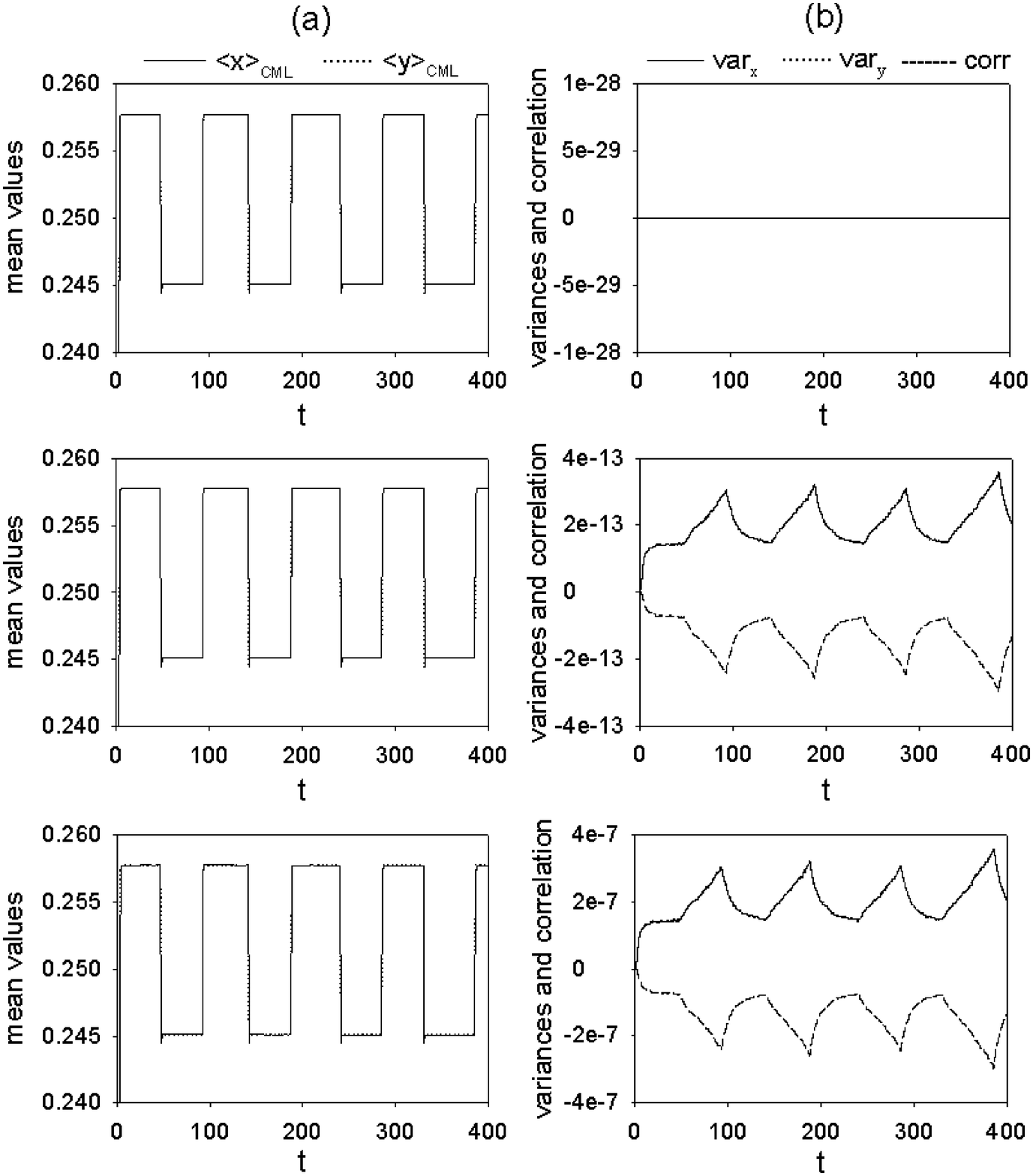}
\end{center}
\vskip-0.6cm \caption{ \small (a) Mean values: $<x>_{_{CML}}$,
$<y>_{_{CML}}$, and (b) variances: $var_x$, $var_y$ and $corr$ of
the two species, as a function of time. The values of the
multiplicative noise intensity are: $\sigma=0, 10^{-12}, 10^{-6}$,
from top to bottom. The initial values of the species
concentrations are $x_{i,j}^{(0)} = y_{i,j}^{(0)} = 0.1$ for all
sites $(i,j)$. The values of the other parameters are the same of
Fig.~\ref{moment_series}.
\bigskip}
 \label{CML_series}
 \vskip-0.4cm
\end{figure}

\vskip-0.2cm In order to check our results we consider a different
approach to analyze the dynamics of our spatial extended system. We
consider the time evolution of CML model, which is the discrete
version of the Lotka-Volterra equations with diffusive terms. For
this model we found anticorrelated spatial patterns of the two
competing species \cite{Valenti1}, that are related to the dynamical
behavior of the moments of the species densities. Here we calculate
the moments in the CML model. Within this formalism, the dynamics of
the spatial distribution of the two species is given by the
following equations
\begin{eqnarray}
x_{i,j}^{(n+1)}&=&\mu x_{i,j}^{(n)} (1-x_{i,j}^{(n)}-\beta^{(n)}
y_{i,j}^{(n)})+\sqrt{\sigma_x} x_{i,j}^{(n)} \xi_{i,j}^{x{(n)}} +
D\sum_\gamma (x_{\gamma}^{(n)}-x_{i,j}^{(n)}),
\label{CLM-Lotka_1}\\
y_{i,j}^{(n+1)}&=&\mu y_{i,j}^{(n)} (1-y_{i,j}^{(n)}-\beta^{(n)}
x_{i,j}^{(n)})+\sqrt{\sigma_y} y_{i,j}^{(n)} \xi_{i,j}^{y{(n)}} +
D\sum_\gamma (y_{\gamma}^{(n)}-y_{i,j}^{(n)}), \label{CLM-Lotka_2}
\end{eqnarray}
where $x^{(n)}_{i,j}$ and $y^{(n)}_{i,j}$ denote respectively the
densities of prey x and prey y in the site $(i,j)$ at the time step
$n$, $\mu$ is the growth rate and $D$ is the diffusion constant.
$\xi_{i,j}^{x(n)}$ and $\xi_{i,j}^{y(n)}$ are independent Gaussian
white noise sources with zero mean and unit variance. The
interaction parameter $\beta^{(n)}$ corresponds to the value of
$\beta(t)$ taken at the time step $n$, according to
Eq.~(\ref{jump_rate}). Here $\sum_\gamma$ indicates the sum over the
four nearest neighbors. To evaluate the $1^{st}$ and $2^{nd}$ order
moments we define on the lattice, at the time step $n$, the mean
values $<x>_{_{CML}}^{(n)}$, $<y>_{_{CML}}^{(n)}$,
\begin{equation}
<z>_{_{CML}}^{(n)}~=~\frac{\sum_{i,j}z^{(n)}_{i,j}}{N}~, \qquad z =
x, y \label{CML moments}
\end{equation}
the variances $var^{(n)}_x$, $var^{(n)}_y$
\begin{equation}
var^{(n)}_z = \sqrt{s^{(n)}_z}=\sqrt{\frac{\sum_{i,j}
(z^{(n)}_{i,j}-<z>^{(n)})^2}{N}}~,  \qquad z = x, y \label{CML
var}
\end{equation}
and the correlation coefficient $corr^{(n)}$ of the two species
\begin{equation}
corr^{(n)} = \frac{cov^{(n)}_{xy}}{s^{(n)}_x s^{(n)}_y},
\label{corr}
\end{equation}
with
\begin{equation}
cov^{(n)}_{xy}=\frac{\sum_{i,j}(x^{(n)}_{i,j}-<x>^{(n)})
(y^{(n)}_{i,j}-<y>^{(n)})}{N}.
 \label{covariance}
\end{equation}
The number of lattice sites is $N = 100\times 100$. The time
behavior of these quantities, for two levels of the multiplicative
noise and in the absence of it, is reported in
Fig.~\ref{CML_series}.
The $1^{st}$ and $2^{nd}$ order moments of Eqs.~(\ref{CML
moments})- (\ref{covariance}) correspond respectively to the same
quantities shown in Fig.~\ref{moment_series}. We note that the two
set of time series are in a good qualitative agreement. The
discrepancies in the oscillation intensities are due to the
different values of the stationary values of the species densities
in the considered models. Specifically: $x_{st} = y_{st} =
\alpha/(1 + \beta) \simeq 0.5$ for the mean field model, and
$x_{st}^n = y_{st}^n = (1 - 1/\mu)/(1 + \beta^n) \simeq 0.25$ for
the CML model. Moreover the behavior of the $2^{nd}$ order moments
in Fig.~\ref{CML_series}b shows little irregularities with respect
to that obtained in the mean field model, because the species
interaction in the CML model is restricted to the nearest
neighbors.

\section{Conclusions}
\vskip-0.2cm We report a study on the dynamics of a spatially
extended ecosystem of two competing species, described by
generalized Lotka-Volterra equations. Two noise sources are present:
a multiplicative white noise, which affects directly the two species
densities, and a correlated dichotomous noise, which produces a
random interaction parameter whose values jump between two levels.
The role of the dichotomous correlated noise is to control the
dynamical regime of the ecosystem (see Fig.~\ref{beta}), while the
multiplicative noise is responsible for the anticorrelated behavior
of the species concentrations (see time behavior of $\mu_{11}$ in
Figs.~\ref{moment_series}b and~\ref{CML_series}b). The
anti-correlated oscillations are enhanced by increasing the
multiplicative noise intensity. The mean field approach with the
Gaussian approximation enables us to obtain the time behavior of the
$1^{st}$ and $2^{nd}$ order moments, which characterize the
spatio-temporal behavior of the ecosystem. We compare the results
obtained within a mean field approach with those obtained with a CML
model. The agreement is quite good and allows us to conclude that
the spatial patterns of the two species, within the mean field
approach, should be non-overlapping as those obtained with the CML
model~\cite{Valenti1}. Our theoretical results could explain the
time evolution of populations in real ecosystems whose dynamics is
strictly dependent on noise sources, which are always present in the
natural environment~\cite{Garcia,Caruso,Sprovieri}.
\section{Acknowledgments}
\vskip-0.2cm This work was supported by ESF (European Science
Foundation) STOCHDYN network and partially by MIUR.


\begin{thebibliography}{99}

\bibitem{Zimmer}
Special section on Complex Systems, Science \textbf{284}, 79
(1999); C. Zimmer, Science, 284 (1999) 83; O. N. Bjornstad and B.
T. Grenfell, Science \textbf{293}, 638 (2001).

\bibitem{Ciuchi}S. Ciuchi, F. de Pasquale and B. Spagnolo,
Phys. Rev. E \textbf{53}, 706 (1996); M. Scheffer \emph{et al.},
Nature \textbf{413}, 591 (2001); A. F. Rozenfeld et al., Phys.
Lett. A \textbf{280}, 45 (2001).

\bibitem{Kaneko}Special issue CML models, edited by K.
Kaneko [Chaos \textbf{2}, 279-- (1992)].

\bibitem{Spagnolo}
B. Spagnolo, D. Valenti, A. Fiasconaro, Math. Biosciences and
Engineering \textbf{1}, 185-211 (2004).

\bibitem{Garcia}J. Garc\'ia Lafuente, A. Garc\'ia, S. Mazzola,
L. Quintanilla, J. Delgado, A. Cuttitta and B. Patti, Fishery
Oceanography \textbf{11}, 31 (2002).

\bibitem{Vilar-Spagnolo}J. M. G. Vilar and R. V. Sol\'e,
Phys. Rev. Lett. \textbf{80}, 4099 (1998); B. Spagnolo and A. La
Barbera, Physica A \textbf{315}, 114-124 (2002); A. La Barbera and
B. Spagnolo, Physica A \textbf{315}, 201 (2002); B. Spagnolo, A.
Fiasconaro, D. Valenti, Fluc. Noise Lett. \textbf{3}, L177 (2003).

\bibitem{Valenti}
D. Valenti, A. Fiasconaro, B. Spagnolo, Mod. Prob. Stat. Phys.
\textbf{2}, 91 (2003); D. Valenti, A. Fiasconaro and B. Spagnolo,
Physica A \textbf{331}, 477 (2004).

\bibitem{Valenti1}D. Valenti, A. Fiasconaro, B. Spagnolo,
Acta Phys. Pol. B \textbf{35}, 1481 (2004).

\bibitem{Benzi}R. Benzi, A. Sutera, A. Vulpiani,
J. Phys.: Math Gen. \textbf{14}, L453 (1981); L. Gammaitoni, P.
Hanggi, P. Jung, and F. Marchesoni, Rev. Mod. Phys. \textbf{70},
223 (1998); V. S. Anishchenko, A. B. Neiman, F. Moss, and L.
Schimansky-Geier, Phys. Usp. \textbf{42}, 7 (1999); T. Wellens, V.
Shatokhin, and A. Buchleitner, Rep. Prog. Phys. \textbf{67}, 45
(2004).

\bibitem{Kawai}R. Kawai, X. Sailer, L. Schimansky-Geier, C. Van den
Broeck, Phys. Rev. E \textbf{69}, 051104 (2004).

\bibitem{Caruso}A. Caruso, M. Sprovieri, A. Bonanno, R. Sprovieri,
Riv. Ital. Paleont. Strat. \textbf{108}, 297 (2002).

\bibitem{Sprovieri}R. Sprovieri, E. Di Stefano, A. Incarbona, M. E. Gargano,
Palaeogeography, Palaeoclimatology, Palaeoecology \textbf{202},
119 (2003).

\end{thebibliography}
\end{document}